\documentclass[twocolumn,aps,pra,reprint,superscriptaddress,longbibliography]{revtex4-1}
\usepackage{amsmath}
\usepackage{graphicx}
\usepackage[lofdepth,lotdepth, caption=false]{subfig}
\usepackage{verbatim}
\usepackage{color}
\usepackage{dcolumn}
\usepackage{bm}
\usepackage{miller}
\usepackage{color}

\usepackage{xr-hyper}
\usepackage{hyperref}

\usepackage[english]{babel}

\begin{document}

\title{Appearance of T$_{d}^{*}$ phase across the T$_{d}$--1T$^{\prime}$ phase boundary in Weyl semimetal MoTe$_{2}$}

\author{Yu Tao}
\affiliation{Department of Physics, University of Virginia, Charlottesville, Virginia 22904, USA}

\author{John A.~Schneeloch}
\affiliation{Department of Physics, University of Virginia, Charlottesville, Virginia 22904, USA}

\author{Chunruo Duan}
\affiliation{Department of Physics, University of Virginia, Charlottesville, Virginia 22904, USA}

\author{Masaaki Matsuda}
\affiliation{Neutron Scattering Division, Oak Ridge National Laboratory, Oak Ridge, Tennessee 37831, USA}

\author{Sachith E.\ Dissanayake}
\thanks{Present address: Duke University, Dept.\ of Physics, Durham, NC 27708}
\affiliation{Neutron Scattering Division, Oak Ridge National Laboratory, Oak Ridge, Tennessee 37831, USA}

\author{Adam A.~Aczel}
\affiliation{Neutron Scattering Division, Oak Ridge National Laboratory, Oak Ridge, Tennessee 37831, USA}
\affiliation{Department of Physics and Astronomy, University of Tennessee, Knoxville, Tennessee 37996, USA}

\author{Jaime A.\ Fernandez-Baca}
\affiliation{Neutron Scattering Division, Oak Ridge National Laboratory, Oak Ridge, Tennessee 37831, USA}

\author{Feng Ye}
\affiliation{Neutron Scattering Division, Oak Ridge National Laboratory, Oak Ridge, Tennessee 37831, USA}

\author{Despina Louca}
\thanks{Corresponding author. Email: louca@virginia.edu}
\affiliation{Department of Physics, University of Virginia, Charlottesville, Virginia 22904, USA}

\begin{abstract}
Using elastic neutron scattering on single crystals of MoTe$_{2}$ and Mo$_{1-x}$W$_{x}$Te$_{2}$ ($x \lesssim 0.01$), the temperature dependence of the recently discovered T$_{d}^{*}$ phase, present between the low temperature orthorhombic T$_{d}$ phase and high temperature monoclinic 1T$^{\prime}$ phase, is explored. The T$_{d}^{*}$ phase appears only on warming from T$_{d}$ and is observed in the hysteresis region prior to the 1T$^{\prime}$ transition. This phase consists of four layers in its unit cell, and is constructed by an ``AABB'' sequence of layer stacking operations rather than the ``AB'' and ``AA'' sequences of the 1T$^{\prime}$ and T$_{d}$ phases, respectively. Though the T$_{d}^{*}$ phase emerges without disorder on warming from T$_{d}$, on cooling from 1T$^{\prime}$ diffuse scattering is observed that suggests a frustrated tendency toward the ``AABB'' stacking. 

\end{abstract}

\maketitle


Many layered materials have structure-property relationships that depend on their layer stacking. For example, the transition metal dichalcogenide MoTe$_{2}$ is reported to be a type-II Weyl semimetal in its orthorhombic T$_{d}$ phase (with the non-centrosymmetric $Pnm2_1$ space group) \cite{sun2015prediction,deng2016experimental}, but not in its monoclinic 1T$^{\prime}$ phase (with the centrosymmetric space group $P2_1/m$). The two phases have nearly-identical layers and differ mainly by  in-plane displacements.
Though there is much interest in investigating Weyl semimetals, the properties of MoTe$_{2}$ are not completely understood. For instance, there is much debate on the origin of the extreme magnetoresistance observed at low temperatures \cite{rhodes_bulk_2017,zhou_hall_2016,thirupathaiah_mote2:_2017}, the number and location of Weyl points in the T$_{d}$ phase \cite{weber_spin-resolved_2018}, and the topological nature of the observed surface Fermi arcs that are a necessary but not sufficient condition for a Weyl semimetal \cite{xu_evidence_2018}. 
Structural distortions have been known to occur, such as stacking disorder during the phase transition, evidenced by the presence of diffuse scattering observed in neutron \cite{schneeloch_emergence_2019} and X-ray \cite{clarke1978low} experiments, and hysteresis effects that extend far beyond the transition region, as seen in resistivity measurements along the thermal hysteresis loop \cite{zandt_quadratic_2007}. These effects have been largely ignored, though one of the surface Fermi arcs was noted to persist to $\sim$90 K above the transition temperature and to have a history-dependent appearance \cite{weber_spin-resolved_2018}. 
In general, structural phase transitions that involve in-plane translations of layers resulting from changes in temperature or pressure have been neglected, but many materials fall in this category, including 
Ta$_{2}$NiSe$_{5}$ \cite{nakano_pressure-induced_2018}, 
In$_{2}$Se$_{3}$ \cite{ke_interlayer-glide-driven_2014, zhao_structure_2014}, 
$\alpha$-RuCl$_{3}$ \cite{glamazda_relation_2017}, 
CrX$_{3}$ (X=Cl, Br, I) \cite{mcguire_coupling_2015}, 
and MoS$_{2}$ \cite{hromadova_structure_2013, chi_pressure-induced_2014, nayak_pressure-induced_2014}. A better understanding of these types of transitions would not only elucidate these material properties, but could also lead to the discovery of new phases. 

\begin{figure}[h]
\begin{center}
\includegraphics[width=8.6cm]
{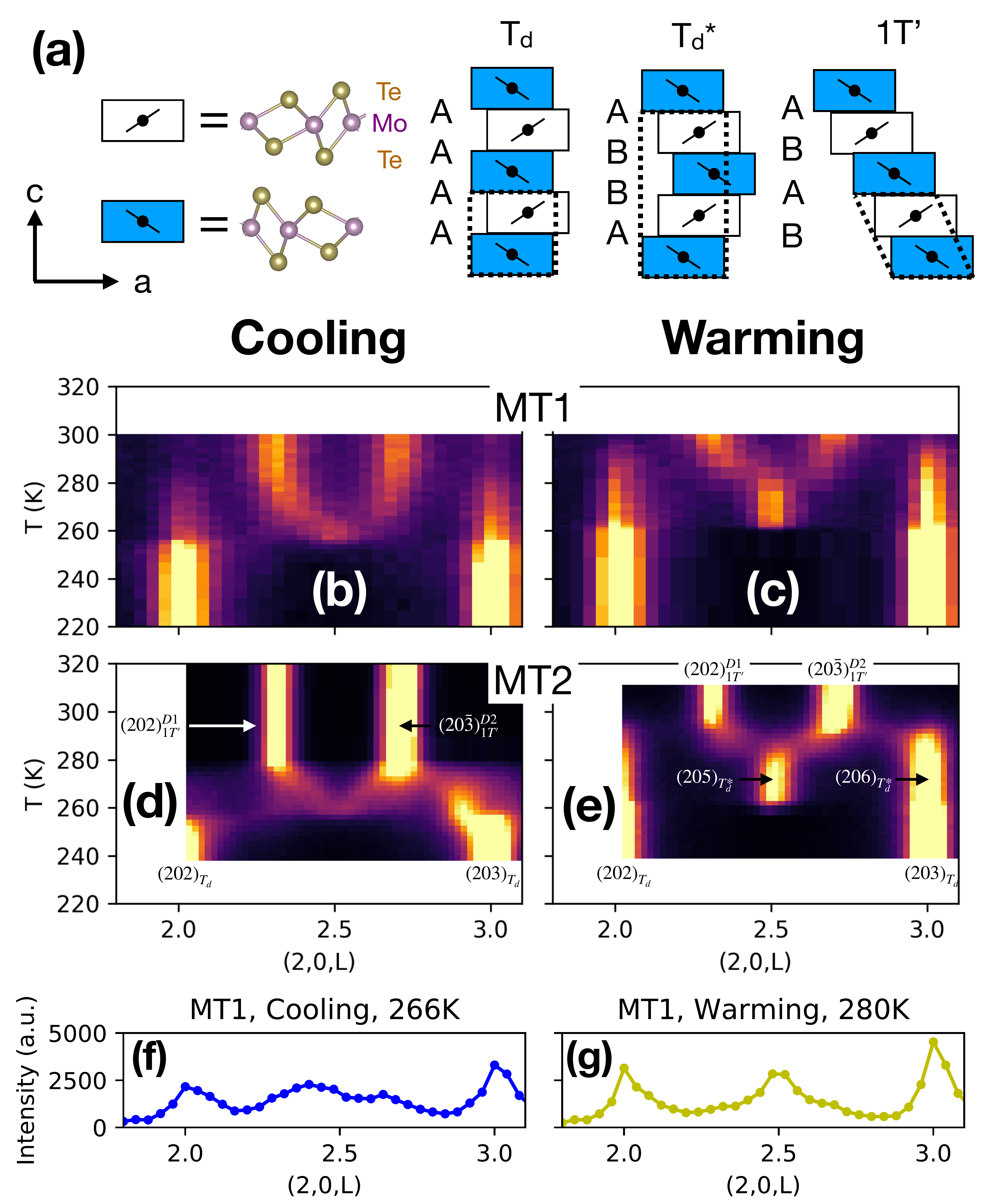}
\end{center}
\caption{(a) The stacking patterns of 1T$^{\prime}$, T$^{*}_{d}$, and T$_{d}$. Rectangles show cells centered on points of inversion symmetry for each layer. Neutron scattering intensity maps for (b,c) MT1 and (d,e) MT2 as a function of temperature along the $(2,0,L)$ line on cooling (left) and warming (right). %
Data were taken on HB1A for (b,c) and HB1 for (d,e). (f,g) Intensity plots along $(2,0,L)$ showing diffuse scattering in MT1 on cooling (f) and warming (g). 
}
\label{fig:Schematic}
\end{figure}

The T$_{d}$ and 1T$^{\prime}$ phases can be constructed from a stacking pattern of ``A'' and ``B'' operations, as shown in Fig.\ \ref{fig:Schematic}(a). The A operation maps one layer of T$_{d}$ to the layer below it, so T$_{d}$ can be built from repeating ``AA'' sequences. The B operation is the same as for A but followed by a translation of $\pm$0.15 lattice units, with a sign that is alternating layer-by-layer. Thus, 1T$^{\prime}$ can be built from repeating ``AB'' sequences. We previously reported that diffuse scattering observed in the $H0L$ scattering plane on cooling from 1T$^{\prime}$ towards T$_{d}$ (in particular, the low intensity along $(60L)$) is consistent with a disordered A/B stacking pattern \cite{schneeloch_emergence_2019}. How the stacking changes with temperature has not been closely examined, though an explanation for the relative stability of the T$_{d}$ and 1T$^{\prime}$ phases via free energy calculations was earlier proposed \cite{kim_origins_2017}. Understanding the nature of layer stacking will provide useful insight into how Weyl nodes disappear across the phase boundary.

We performed elastic neutron scattering as a function of temperature to study the mechanism of the structural phase transition between the 1T$^{\prime}$ and T$_{d}$ phases in MoTe$_{2}$. On warming, the recently discovered T$_{d}^{*}$ phase \cite{dissanayake_electronic_2019} was observed, having a pseudo-orthorhombic structure and a four-layer unit cell, rather than the two-layer unit cells of 1T$^{\prime}$ and T$_{d}$. 
The stacking sequence of T$_{d}^{*}$ can be described by ``AABB'', as shown in Fig. \ref{fig:Schematic}(a). Upon warming, the T$_{d}$$\rightarrow$T$_{d}^{*}$ transition is not accompanied by disorder. Diffuse scattering is observed on further warming from T$^{*}_{d}$ to 1T$^{\prime}$. On the other hand, on cooling from 1T$^{\prime}$ to T$_{d}$, the T$_{d}^{*}$ phase is absent and only diffuse scattering is observed that suggests a frustrated tendency toward the ``AABB'' layer order.


Elastic neutron scattering was performed at Oak Ridge National Laboratory, on the triple axis spectrometers HB1, CG4C, and HB1A at the High Flux Isotope Reactor; and on the time-of-flight spectrometer CORELLI at the Spallation Neutron Source \cite{ye_implementation_2018}. Though the crystals are monoclinic at room temperature, for simplicity, we use orthorhombic coordinates, with $a \approx 6.3$ \AA\, $b \approx 3.5$ \AA\, and $c \approx 13.8$ \AA. 
The collimations were 48$^{\prime}$-40$^{\prime}$-S-40$^{\prime}$-120$^{\prime}$ for HB1 and CG4C, and 40$^{\prime}$-40$^{\prime}$-S-40$^{\prime}$-80$^{\prime}$ for HB1A. Incident neutron energies were 13.5 meV for HB1, 4.5 meV for CG4C, and 14.6 meV for HB1A. Resistance measurements were performed in a Quantum Design Physical Property Measurement System. All crystals were grown in excess Te flux, including the two used for neutron scattering, ``MT1'' and ``MT2''. MT1 has the composition MoTe$_{2}$, while MT2 has the composition Mo$_{1-x}$W$_{x}$Te$_{2}$ with $x \lesssim 0.01$ as estimated by energy dispersive X-ray spectroscopy and the $c$-axis lattice constant. Details can be found in the Supplemental Materials.



In Figures \ref{fig:Schematic}(b-e), neutron scattering intensity scans along $(2,0,L)$ are combined for many temperatures on cooling and warming through the hysteresis. In the 1T$^{\prime}$ phase, the $(202)_{1T^{\prime}}^{D1}$ and $(20\bar{3})_{1T^{\prime}}^{D2}$ Bragg peaks are observed near $L=2.3$ and $L=2.7$, respectively; $D1$ and $D2$ denote each of the two 1T$^{\prime}$ twins. (Since MT1 could not be warmed fully into 1T$^{\prime}$ in Fig.\ \ref{fig:Schematic}(c) for technical reasons, diffuse scattering was present on subsequent cooling from 300 K in Fig.\ \ref{fig:Schematic}(b).)
At low temperatures, T$_{d}$-phase Bragg peaks at $L=2$ and $L=3$ are observed, as indicated in the figure. On warming from T$_{d}$ past $\sim$260 K, a peak appears at L = 2.5, indicating the onset of T$_{d}^{*}$. 
The presence of this peak at half-integer $L$ indicates an out-of-plane doubling of the unit cell, so we label this peak $(205)_{T_{d}^{*}}$ (Fig.\ \ref{fig:Schematic}(e)). With additional warming, a gradual transformation into the 1T$^{\prime}$ phase occurs, accompanied by diffuse scattering indicating stacking disorder. 
Examples of the diffuse scattering can be seen in the individual plots of intensity along $(2,0,L)$ in Fig.\ \ref{fig:Schematic}(f), where 1T$^{\prime}$ is transitioning into T$_{d}$;  and in Fig.\ \ref{fig:Schematic}(g), where T$_{d}^{*}$ is transitioning into 1T$^{\prime}$. 
For MT2, we measured through the hysteresis twice and found the same pattern of diffuse scattering at the same temperatures along the hysteresis, suggesting that the appearance of the diffuse scattering through the hysteresis is reproducible.

\begin{figure}[h]
\begin{center}
\includegraphics[width=8.6cm]
{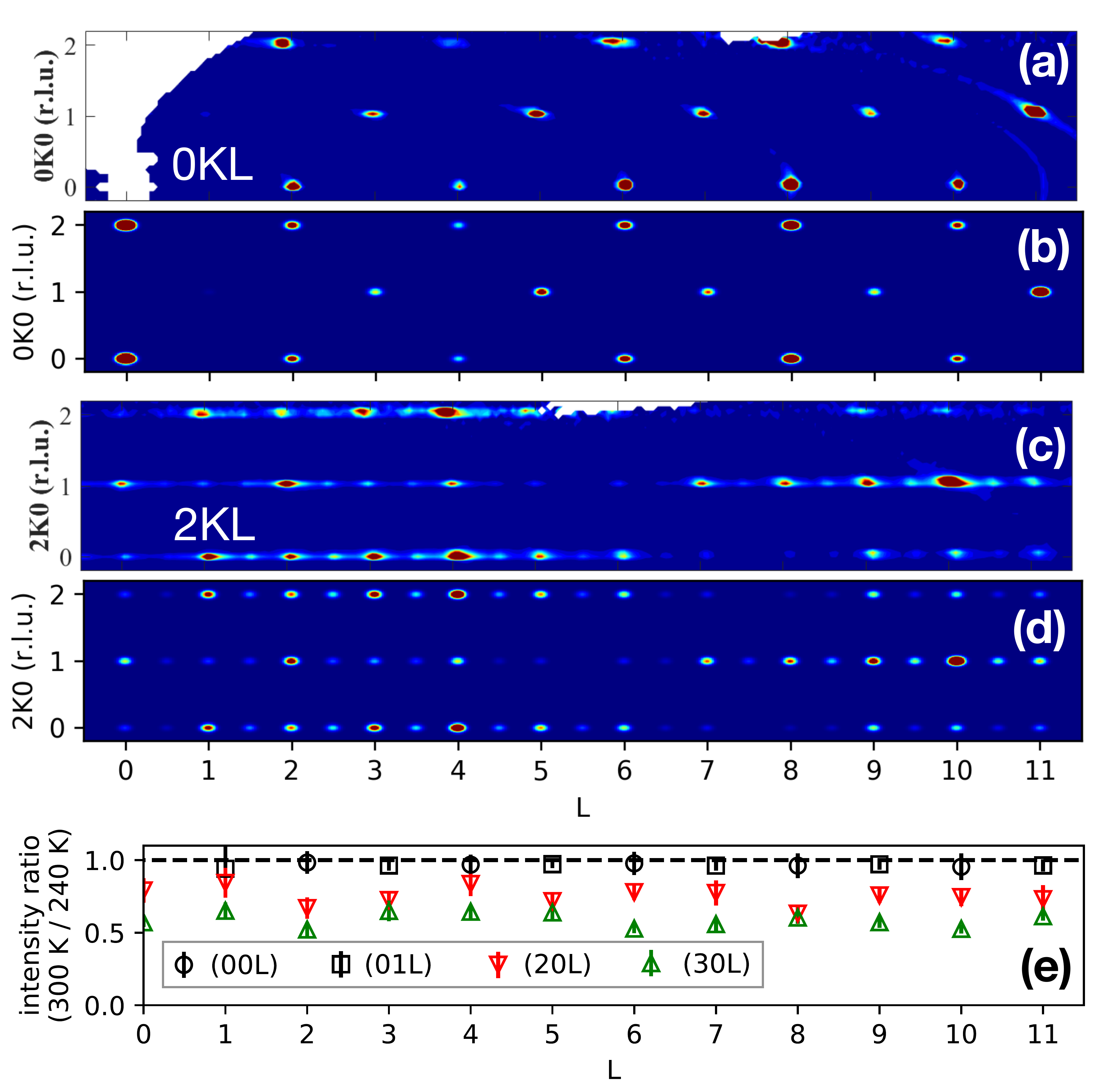}
\end{center}
\caption{Neutron scattering intensity maps (a,c) and simulated data (b,d) in the $0KL$ (a,b) and $2KL$ (c,d) scattering planes in the T$_{d}^{*}$ phase, from the MT1 crystal measured at CORELLI. Data taken on warming at 300 K. (e) Ratio of selected Bragg peak intensities between 300 K and 240 K. Intensities are from Gaussian fits from data averaged within $\pm$0.2 r.l.u.\ in the $H$- and $K$-directions.}
\label{fig:CORELLI}
\end{figure}

The T$_{d}^{*}$ phase structure can be deduced from the following observations: First, T$_{d}^{*}$ appears to be orthorhombic, but has additional peaks at half-integer $L$ values relative to T$_{d}$, indicating a four-layer unit cell. Second, the 1T$^{\prime}$ and T$_{d}$ phases can be built from A/B stacking sequences, so we presume the same is true for T$_{d}^{*}$. There are only two possible pseudo-orthorhombic stacking sequences:  ``AABB'' and ``ABBA'', which are twins of each other. Since this structure, with highest possible symmetry P2$_{1}$/m, appears to have an orthorhombic unit cell but has atomic positions incompatible with orthorhombic space groups, we refer to it as pseudo-orthorhombic.

To verify the predicted AABB stacking structure of T$_{d}^{*}$, we carried out single-crystal neutron diffraction measurements on the MT1 crystal on CORELLI, and the data are shown in Figures \ref{fig:CORELLI}(a,c). The data were taken on warming to 300 K, and the presence of peaks at half-integer $L$ in the $2KL$ plane in Fig.\ \ref{fig:CORELLI}(c) confirm the presence of the T$_{d}^{*}$ phase. 
The diffuse scattering streaks along $L$ are from stacking disorder that was already present on warming from 240 K, possibly due to not cooling sufficiently into T$_{d}$ beforehand.
(There is a discrepancy between the detection of T$_{d}^{*}$ in MT1 at 300 K on CORELLI and up to $\sim$280 K on HB1A. The cause of the temperature discrepancy is unknown, but may be related to the presence of stacking disorder before warming.)
Figures \ref{fig:CORELLI}(b,d) show simulated intensity maps. To match the data, it was necessary to consider a 47.8\% volume fraction of T$_{d}$ as well as 28.2\% and 24.0\% volume fractions of the two T$_{d}^{*}$ twins. The volume fractions were obtained by fitting the intensities of Bragg peaks within $-1\leq H\leq8$, $-1\leq K\leq1$, and $-20\leq L\leq20$ with the calculated peak intensities of the ideal ``AA'', ``AABB'', and ``ABBA'' stacking sequences of T$_d$ and the two T$_d^*$ twins, respectively. These structures were built from layers having the coordinates in Ref.\ \cite{qi_superconductivity_2016}. As can be seen in Figures \ref{fig:CORELLI}(a-d), the patterns of peak intensities in these scattering planes match those arising from our model. 

Stringent constraints on possible T$_{d}^{*}$ structures follow from the lack of change in $(00L)$ and $(01L)$ peak intensities between the T$_{d}$ phase at 240 K and the T$_{d}^{*}$ phase at 300 K (as seen from the near-unity intensity ratios in Fig.\ \ref{fig:CORELLI}(e). For context, intensity ratios for $(20L)$ and $(30L)$ peaks are also included.) A lack of change in $0KL$ peak intensities implies a lack of change in atomic positions along the $b$- and $c$-directions, but is consistent with layer displacements along the $a$-direction, as is the case between 1T$^{\prime}$ and T$_{d}$ \cite{schneeloch_emergence_2019}. The AABB structure should be centrosymmetric, since it can be transformed from the centrosymmetric AB-stacked 1T$^{\prime}$ phase by a centrosymmetric series of translations (see Supplementary Materials). Inversion symmetry centers for the AABB structure are depicted in Fig.\ S2 in the Supplemental Materials. Barring small non-centrosymmetric distortions, which are unlikely given that first-principles calculations have shown that MoTe$_{2}$ layers isolated from the non-centrosymmetric T$_{d}$ environment tend to become centrosymmetric \cite{heikes_mechanical_2018}, we conclude that T$_{d}^{*}$ is centrosymmetric with P$2_{1}/m$ symmetry. A structural refinement assuming P$2_1/m$ symmetry was performed (see Supplementary Materials), with rough agreement between the refined and ideal coordinates, though the absence of visible $0KL$ peaks in our data (apart from those with even $K+L$) indicates that the true T$_d^*$ structure is closer to the ideal AABB stacking than our refined coordinates.


\begin{figure}[h]
\begin{center}
\includegraphics[width=8.6cm]
{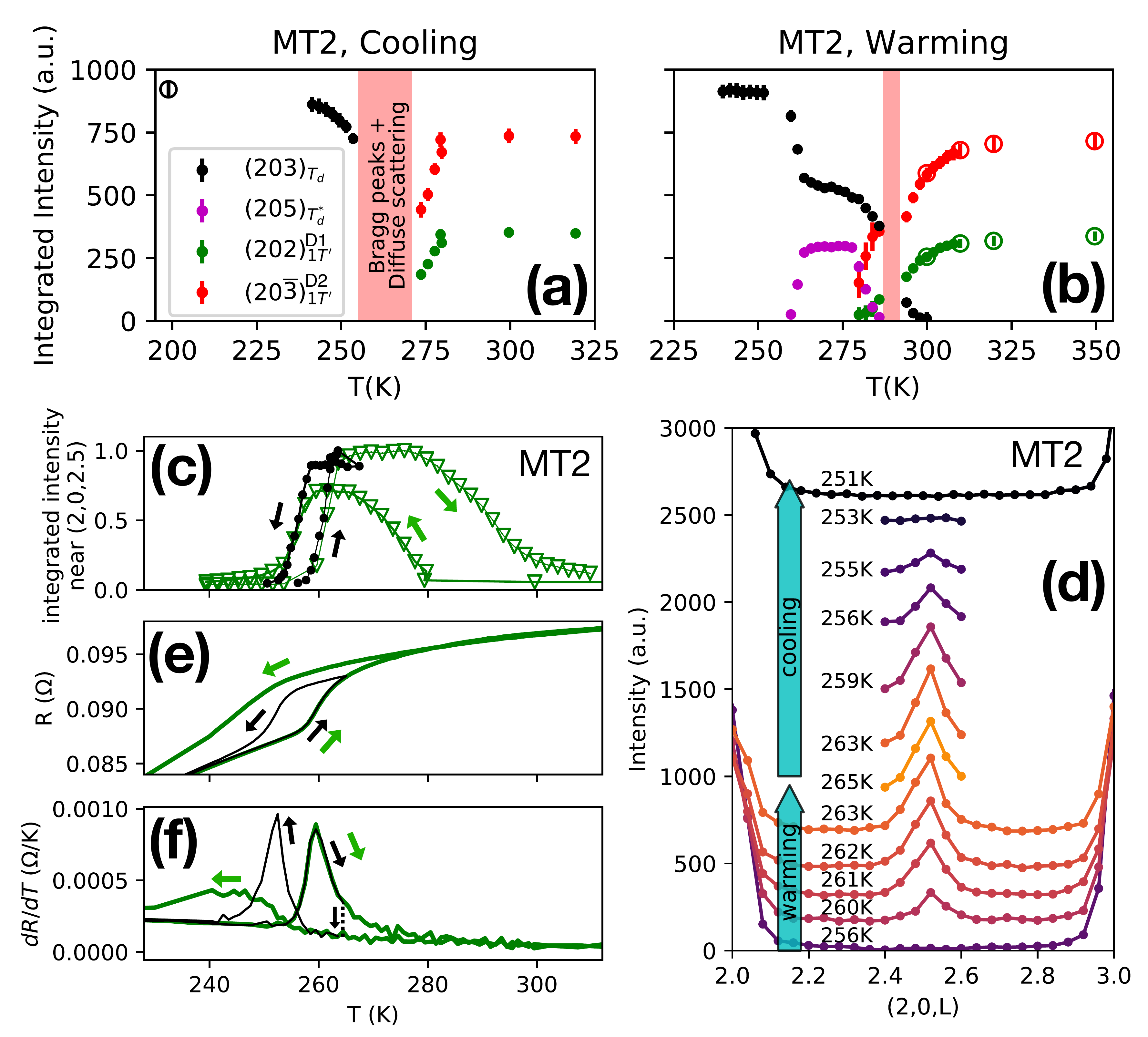}
\end{center}
\caption{(a,b) Bragg peak intensities plotted as a function of temperature on warming and cooling for MT2. %
Red bands denote regions where fitting was poor. Closed symbols denote fits to the same hysteresis loop (with cooling data measured before warming). Open symbols correspond to a previous hysteresis loop. (c) Plots of intensity integrated within (2, 0, $2.39 \leq L \leq 2.61$) for MT2 taken through two different hysteresis loops. Data taken on CG4C for the narrow hysteresis (black), and on HB1 for the wide hysteresis (green; from same data as Fig.\ \ref{fig:Schematic}(d,e).) Curves normalized to their largest values. (d) Neutron scattering intensity along $(2,0,L)$ for MT2, with data taken on CG4C at various temperatures, vertically displaced for clarity. (e) Resistance of a MoTe$_{2}$ crystal, measured through two hysteresis loops that begin on warming from 200 K. 
(f) The derivative $dR/dT$ of the data shown in (e).}
\label{fig:MT2I}
\end{figure}

For a closer look at how the transition proceeds, in Fig.\ \ref{fig:MT2I}(a,b) we plot intensities for four Bragg peaks as a function of temperature. The integrated intensities were obtained from fits of the data shown in Figures \ref{fig:Schematic}(d) and \ref{fig:Schematic}(e). On cooling below $\sim$280 K (Fig. 3(a)), there is a steady decrease in the intensity of the 1T$^{\prime}$ peaks in the hysteresis region. The peaks eventually become difficult to resolve from the diffuse scattering, and fitting was not done within the region indicated by the pink shaded bar. On further cooling, the T$_{d}$ peaks appear. 

In contrast, on warming the intensities of the T$_{d}$ peaks in Fig.\ \ref{fig:MT2I}(b) remain constant until a sudden change is observed around 260 K. At this temperature, the T$_{d}^{*}$ peaks appear (magenta symbols) at the expense of the T$_{d}$ peaks. On further warming, the T$_{d}$ and T$_{d}^{*}$ peak intensities both decrease and disappear by 280 K. Again, diffuse scattering is observed (pink shaded region) prior to the crystal transforming fully into 1T$^{\prime}$. Though a coherent, long-range T$_{d}^{*}$ phase only appears on warming, the intensity shift toward $(2,0,2.5)$ on cooling as seen in Fig.\ \ref{fig:Schematic}(b) and \ref{fig:Schematic}(d) suggests a tendency toward the AABB stacking, though frustrated and not resulting in an ordered structure. 
On both cooling into T$_{d}$ or warming into 1T$^{\prime}$, there is a gradual increase in the intensity of the T$_{d}$ and 1T$^{\prime}$ peaks, which occurs with a decrease in diffuse scattering (see Fig.\ S3 of the Supplemental Materials). 
This lingering diffuse scattering is probably related to the long residual hysteresis commonly observed in the resistivity measurements (e.g., in Ref.\ \cite{zandt2007quadratic}, or Fig.\ S4 in the Supplemental Materials.)

To investigate the boundary between the T$_{d}$ and T$_{d}^{*}$ regions, in Fig.\ \ref{fig:MT2I}(c), intensity integrated near $(2,0,2.5)$ is plotted for two different thermal hysteresis loops for the MT2 crystal. 
The narrow hysteresis loop (black symbols) corresponds to the sample warming into T$_{d}^{*}$, then cooling back to T$_{d}$ without entering 1T$^{\prime}$. Fig.\ \ref{fig:MT2I}(d) is a plot of the data used to calculate the narrow hysteresis loop intensities. The $(205)_{T_{d}^{*}}$ peak intensity rises and falls through the hysteresis loop. Diffuse scattering is not present even at a temperature a few Kelvin below the T$_{d}^{*}$ peak's disappearance on cooling. Thus, T$_{d}^{*}$$\rightarrow$T$_{d}$ likely proceeds without disorder. 
In contrast, a wider hysteresis loop (green symbols) is observed when the sample is allowed to warm into 1T$^{\prime}$. This is coupled to the substantial diffuse scattering present on cooling, as shown earlier in Fig.\ \ref{fig:Schematic}(d,e). Nevertheless, for both narrow and wide hysteresis loops, a sudden drop of intensity near $(2,0,2.5)$ appears on cooling below 255 K, though more gradually for the wide hysteresis loop.

A similar pattern can be seen in the resistance data of Fig.\ \ref{fig:MT2I}(e), taken on a MoTe$_{2}$ crystal with residual resistance ratio $\sim$460 through consecutive narrow (black; 200 to 265 K) and wide (green; 200 to 350 K) hysteresis loops. On cooling, the resistance decreases quickly and in a symmetric manner (for cooling vs.\ warming) for the narrow hysteresis loop, but more slowly and asymmetric for the wide hysteresis loop. Even so, the temperature at which both loops begin to bend on cooling is similar, as seen from $dR/dT$ in Fig.\ \ref{fig:MT2I}(f), though slightly lower for the wide hysteresis loop. The kink seen on warming (near 258 K) is likely the onset of T$_{d}^{*}$ and not 1T$^{\prime}$, judging from the temperature and the similarities between the resistance and neutron scattering hysteresis loops.


We next discuss how these structural transitions proceed and the kinds of interlayer interactions that may be responsible, beginning from the observation that the onset to T$_{d}$ occurs at a similar temperature whether cooling from the ordered T$_{d}^{*}$ phase, or from the frustrated T$_{d}^{*}$ region accessed on cooling from 1T$^{\prime}$. Since the onset temperature to T$_{d}$ does not appear to vary substantially with overall stacking disorder, we suggest that \emph{short-range} rather than long-range interlayer interactions determine the onset temperatures (into 1T$^{\prime}$ or T$_{d}^{*}$ as well as T$_{d}$). 
(Though we use the term ``interlayer interactions'', we emphasize that these are effective interactions. Whether an interlayer boundary shifts from A$\rightarrow$B depends on the free energy, which depends on the surrounding environment, which is specified by the A/B stacking sequence. ``Interlayer interactions'' represent the dependence of an interlayer boundary's contribution to the free energy on the surrounding stacking, and can be indirect, involving changes to band structure, phonon dispersion, etc.)

In contrast, \emph{long-range} interlayer interactions may govern the gradual decrease in diffuse scattering and increase in Bragg peak intensities on warming into 1T$^{\prime}$ or cooling into T$_{d}$. What kind of stacking faults causing this diffuse scattering persist on cooling into T$_{d}$, even when short-range interlayer interactions favor an ordered phase? At twin boundaries, shifts of A$\rightarrow$B or B$\rightarrow$A (e.g., AAA\textbf{A}BBB...$\rightarrow$AAA\textbf{B}BBB...) would not change the short-range environment, and could only be induced by changes in long-range interlayer interactions. The decrease in diffuse scattering in T$_{d}$ on cooling can be explained by the annihilation of these twin boundaries, either by joining in pairs or by exiting a crystal surface. The lack of change on subsequent warming can be explained by the relaxation of conditions that, on cooling, had driven twin boundaries to annihilate.

Previous studies on MoTe$_{2}$ should be re-examined in light of the existence of the T$_{d}^{*}$ phase. First, the hysteresis loop in resistivity (first reported in Ref.\ \cite{hughes_electrical_1978}) has been interpreted as indicating the transition between T$_{d}$ and 1T$^{\prime}$, but in view of the current data, most of the change in the resistance occurs between T$_{d}$ and T$_{d}^{*}$ on warming. 
Second, second harmonic generation (SHG) intensity measurements, expected to be zero for inversion symmetry and nonzero otherwise, show abrupt (within $<$4 K) changes on both heating and cooling through the hysteresis loop \cite{sakai_critical_2016}. Since the transition to 1T$^{\prime}$ occurs gradually, and since T$_{d}^{*}$ appears to be centrosymmetric, the abrupt warming transition seen in SHG may be due to the T$_{d}$$\rightarrow$T$_{d}^{*}$ rather than T$_{d}$$\rightarrow$1T$^{\prime}$ transition. The abrupt transition on cooling is harder to explain, but it is possible that the loss of inversion symmetry on cooling into T$_{d}$ occurs suddenly even as the transition proceeds with disorder. 
Our findings may also inform proposed applications, such as the photoinduced ultrafast topological switch in Ref.\ \cite{zhang_light-induced_2019}; since the T$_{d}$$\rightarrow$T$_{d}^{*}$$\rightarrow$T$_{d}$ transition occurs without disorder and with only a $\sim$5 K hysteresis, and since T$_{d}^{*}$ appears to be centrosymmetric, a topological switch may more efficiently use T$_{d}^{*}$ rather than 1T$^{\prime}$. 
%

In conclusion, using elastic neutron scattering, we mapped the changes in stacking that occur in the thermal hysteresis between the T$_{d}$ and 1T$^{\prime}$ phases in MoTe$_{2}$. On warming from the orthorhombic T$_{d}$, T$_{d}^{*}$ arises without diffuse scattering and corresponds to an ``AABB'' sequence of stacking operations. 
Diffuse scattering is present on further warming from T$_{d}^{*}$ to 1T$^{\prime}$, and on cooling from 1T$^{\prime}$ to T$_{d}$, where a frustrated tendency toward the ``AABB'' stacking is seen. 
Thus, the 1T$^{\prime}$-T$_{d}$ transition has complex structural behavior and deserves further study.

\section*{Acknowledgements}

This work has been supported by the Department of Energy, Grant number
DE-FG02-01ER45927.  A portion of this research used resources at the High Flux Isotope Reactor and the Spallation Neutron Source, which are DOE Office of Science User Facilities operated by Oak Ridge National Laboratory.

\end{document}